\newcommand{\beq}{\begin{equation}}
\newcommand{\eeq}{\end{equation}}
\newcommand{\beeq}{\begin{eqnarray}}
\newcommand{\eeeq}{\end{eqnarray}}

\documentstyle[12pt]{article}
\begin{document}
\author{V.A. \ Kudryavtsev  \\ %EndAName
{\em Petersburg Nuclear Physics Institute,}\\{\em \ 188350,
Gatchina, Russia.}}
\title{{\bf Fermion dominated composite superstring  model and
unified description of hadron and lepton degrees of freedom in it}}
\maketitle

\begin{abstract}
 New string dynamics is formulated on the basis of the extended set
of supergauge constraints including not only supergauge Virasoro
conditions but also nilpotent supercurrent constraints .  This
approach arises from  a natural generalization of the
classical operator many-string vertices.  The formulation of this model
leads to three two-dimensional surfaces for description of hadron
strings. It gives some dynamical generalization of Chan-Paton factor
for string amplitudes in terms of operator vertices.
Supersymmetry on the 2-d world  surface is present but ten-dimensional
supersymmetry is absent.
 In this approach two-dimensional fermion string fields make it possible
to give a unified description of hadron and lepton degrees of freedom
and of its dynamics. This model allows to solve the problem of
elimination of the most part of parity twins in the baryon spectrum.

  One-loop (and many-loops perhaps) amplitudes in this model are finite
due to the extended set of supergauge constraints and to
the significant excess of the total number of fermion two-dimensional
fields over the number of boson 2-d fields.

\end{abstract}

\newpage

\section{Introduction}

%%%

  In spite of significant success in treatment of
 hadron interactions of high energy the quantum chromodynamics
(QCD) up to now is unable to give a consistent quantitative description
of most strong interactions up to 2-5 GeV, that is to say
  of  soft hadron interactions.  Chiral models
\cite{1,2} are enough effective for small energies of light hadrons
$( < 300 MeV)$.  However the chiral approach
solely is incapable to describe the resonance region of hadron
interactions in a consistent way.

  Attempts to build string-based hadron amplitudes have
been undertaken at dawn of the age of string theory (dual resonance
models) \cite{3}. They were initiated by the remarkable universal
linearity of the Regge trajectories for spins of meson and baryon
resonances $J=\alpha(M^2)=\alpha(0)+\alpha^{\prime}_H
M^2$;$\;\;(\alpha^{\prime}_H\approx 0,85 GeV^{-2})$.  Now we have
these trajectories up to $J=5$. From this point of view the $\rho$
-trajectory with $\alpha(0)=1/2$ turns out to be the leading hadron
trajectory. Up to now there is no other satisfactory explanation of the
striking linearity of hadron Regge trajectories than a stringlike
form of hadrons.  This hadron string approach was very encouraging with
respect to the few particles amplitudes $ (\pi\pi\to\pi\pi $ for
example ) in good agreement with experimental data and the chiral
limit.  But many-particles amplitudes and loop corrections found then
proved to break down unitarity.  All further consistent string
amplitudes for many particles have to include massless vector mesons in
the resonance spectrum in order for unitarity requirements to be
satisfied.  They correspond to the $\rho$ -trajectory with
$\alpha(0)=1$.  It is in obvious contradiction with the hadron spectrum
observed experimentally.  Development of this string approach with the
leading Regge trajectory $\alpha(M^2)=J=1+\alpha^{\prime}M^2$ has led
to the usual superstring theory \cite {4} for  interactions of quarks,
gluons and gravitons at Planck distances
$(\alpha^{\prime}\equiv\alpha^{\prime}_P\sim 10^{-38} GeV^{-2})$ beyond
reach of experiment.

   The initial case with the  $\rho$-trajectory
  $\alpha(M^2)=J=1/2+\alpha^{\prime}_H M^2$ was realized by Lovelace
and Shapiro in the amplitude for $\pi\pi\to\pi\pi$ \cite {5}.  It
  corresponds to the correct chiral limit and to the Adler- Weinberg
 condition \cite {6}.  A few years ago a consistent generalization of
Lovelace-Shapiro amplitudes for arbitrary number of pions with the
leading trajectory $\alpha(M^2)=J=1/2+\alpha^{\prime}_H M^2$ has been
  found \cite {7,8}.  New many-pions amplitudes do not contradict
unitarity.

  Second difficulty of the string approach to hadron interactions was to
describe hadron scattering at large angles for high energies s and high
transfer momenta t. It may appear that there is an exponential decrease
for string amplitudes here in obvious contradiction with
experimental data and the parton picture. However  discovery of
sister traectories \cite {9} has proved this exponential falloff
to be reduced up to a power dependence owing to contributions of
many particles states. Moreover as we shall see further both
fundamental values of the slope $\alpha^{\prime}$: the hadronic
   $\alpha^{\prime}_H$
and
the extreme small Planck $\alpha^{\prime}_P$ for closed strings are
possible in this model. Thus this obstacle was been overcome and this
argument against string interpretation of hadrons  has lost its
strength too.

 So the main problem for the string approach to hadrons is to combine
requirements of unitarity and  the intercept of the leading hadron
trajectory to be equal one half (instead of the traditional value to be
  equal one in the case of classical superstring theories).  The
 absence of ghosts (states with negative norm) in the spectrum of
 physical states in traditional string models is provided by the
 Virasoro supergauge constraints to be satisfied.  In turn the
 fulfillment of the Virasoro conditions usually leads to the intercept
 of the leading Regge trajectory for mesons to be equal one and the
 presence of massless vector mesons in the resonance spectrum
 correspondingly.  A possible solution of this problem on the basis of
 some natural generalization of many-string operator vertices (N-reggeon
vertices \cite {10}) was suggested in \cite {8}.  In this model
the physical hadron spectrum does not consist massless vector states
and the leading Regge trajectory for mesons has the intercept to be
equal one half.  It proves to be possible due to a new functional
approach to string amplitudes.  This construction includes some cyclic
functional integration which generalizes traditional one and generates
natural structure of the Harari-Rosner quark dual diagrams \cite {11}
for string amplitudes.  Second new attribute of this model is nilpotent
supercurrent gauge constraints apart from usual Virasoro supergauge
 ones. The previous formulation of this approach in \cite {7} is not
sufficiently advanced to deduce consistent supercurrent constraints and
vertices which will satisfy requirements of unitarity.  A new
modification to be considered here allows to obtain necessary
constraints and operator vertices.  Furthermore two-dimensional fields
of this model are capable of describing all hadron and lepton degrees
of freedom and its interactions.

\section{Fermion two-dimensional fields, functional formulation of the
model and Virasoro supergenerators }

   Two-dimensional fields of the model are represented by two sets.
First set includes usual fields of the Neveu-Schwarz model \cite {12}:
the string coordinate  $X_{\mu}(\xi)$ and its fermion
superpartner $H_{\mu}(\xi)$, $\mu=0,\ldots,d-1$, $d=10$;  $\xi$ are
 coordinates on two-dimensional world surface.  Second set
includes new two-dimensional fermion fields: $\Psi_{\alpha}(\xi)$
and $\Phi_{a}(\xi)$.  New fields $\Psi_{\alpha}$ $\alpha=1,\ldots32$
are spinor fields in ten-dimensional target space (compare with the
 Bardakci-Halpern model \cite {13,14}).  Other fields $\Phi_{a}$ are
two-dimensional fermion fields which carry quantum numbers of
all currents $J_{a}$  composed from $\Psi_{\alpha}$:
 $J_{a}^{(\psi)}=\tilde\Psi \Gamma_{a}\Psi$.  It is easy to find the
 number of independent matrices $ \Gamma_{a}$ and of corresponding
 fields $\Phi_{a}$. We obtain this number to be equal $32*31/2=496$
due to the anticommutation of $\Psi_{\alpha}$ fields. Actually this
 model includes two sets of new fields as we shall see below.
New two-dimensional fermion fields provide a way of obtaining
the following results:

1)   unified description of all quark flavours and lepton degrees of
freedom;

2)   consistent from viewpoint of unitarity   tree
amplitudes for N hadrons due to a new operator construction
of these amplitudes;

3)    finite one loop  (and many loops perhaps) corrections  for
open strings due to the significant excess of the number of fermion
two-dimensional fields over the number of boson fields (including BRST
 fields).

   Let us remind that  the divergence of an amplitude $A_1$ for one
loop planar N-particle diagram in the sector of open strings is defined
by integral over the variable $x=x_{1}x_{2}\dots x_{N}$ of $f(x)$  near
$x=1$.  For $x\rightarrow 1$  this function behaves in the following
  manner:

   $f(x)\rightarrow r \exp {c(D-D')/(1-x)}$.

    Here $D$ is the
 number of boson fields, $D'$ is the number of fermion two-dimensional
 fields; \\
$x_{i}$ are the Chan variables; $\;\; c=\frac{\pi^2}{6}$ ;\\ $r(x)$ is
some power function for $x\rightarrow 1$.  We have   $D=D'$ for usual
 superstring approach and $D'-D=\frac{2^5*(2^5+1)}{2}=512$ for the
    model under consideration. \\

   It is worth  attention that an  nonzero value of the
 one loop amplitude to be given by integral to be considered may be
anomal small in this model due to the factor $r \exp {c(D-D')/(1-x)} $.

    Now we define relations between the real quantum numbers and
 the components of the $\Psi$ and $\Phi$ fields. It is convenient to
 represent 32 components of $\Psi$ as $\Psi_{\alpha\beta\gamma\delta}$
 components. Here $\alpha=1,2,3,4$  is the usual Dirac index;
 $\beta=1,2; \gamma=1,2;\delta= 1,2$ are three isotopic indices which
 are corresponding to internal quantum numbers and give us eight
isotopic components.

   We suppose that six additional dimensions  form  a flat six-dimensional
compact space with enough small sizes and with global asymmetry which
corresponds to the real world.

  It is suitable to
build  $32\otimes32$ matrices in this model in accordance with the
$\Psi_{\alpha\beta\gamma\delta}$ components as direct product of the
$4\otimes4$ Dirac matrices in the Majorana representation and the
   $8\otimes8$ isotopic matrices.

  For instance
$$
   \Gamma_{\mu}=\gamma_{\mu}\otimes I\otimes I\otimes I;
$$
$$
T_{3}^{(1)}=\frac{1}{2}I\otimes \tau_{3}\otimes I\otimes I ;
T_{3}^{(2)}=\frac{1}{2}I\otimes I\otimes \tau_{3}\otimes I ;
T_{3}^{(3)}=\frac{1}{2}I\otimes I\otimes I\otimes \tau_{3} .
$$
Four $\alpha$ components of $\Psi$ carry usual spin  1/2, and eight
  $\beta,\gamma,\delta$ components correspond to the quark flavours
 and the lepton numbers according to the Table 1.  \begin{center}
	   \begin{tabular} {|c|c|c|c|} \hline
	   &\multicolumn{3}{c|}{eigen values}
\\ \cline{2-4}
   flavours&$T_{3}^{(1)}$&$T_{3}^{(2)}$&$T_{3}^{(3)}$ \\
\hline\hline $\nu$ &$ +1/2$ &$ +1/2$ &$ +1/2$ \\ \hline $ e^{-}$ &$
	    -1/2$ &$ -1/2$ &$ -1/2$ \\ \hline $u$   &$ +1/2$ &$ -1/2$
	    &$ -1/2$ \\ \hline $d$ &$ -1/2$ &$ +1/2$ &$ +1/2$ \\ \hline
	   $c$   &$ -1/2 $&$ +1/2$ &$ -1/2$ \\ \hline
	   $s$   &$ +1/2$ &$ -1/2$ &$ +1/2$ \\ \hline
	   $t$ &$ -1/2$ &$ -1/2$ &$ +1/2$ \\ \hline
	   $b$   &$ +1/2$ &$ +1/2$ &$ -1/2$ \\ \hline
\end{tabular}
\end{center}
\begin{center}
              Table I.
   Quantum numbers of the $\Psi$ components
\end{center}

   Absence  of transitions betweeen quark and lepton
degrees of freedom  and the differences in the dynamics of its
interactions at the experiment one can to present as conseqence of
some global asymmetry of the compact target  space and fields on it.
This is a asymmetry  in reference to $T_{3}^{(1)}, T_{3}^{(2)},
T_{3}^{(3)}$  between  "symmetrical" lepton components and
"nonsymmetrical" quark ones.

  Here we shall give the functional definition of the tree
amplitudes of this approach. This representation  generalizes the usual
functional approach and proves to be the direct conseqence of
generalization of many-string operator vertices in papers \cite {7,8}.

  At first let us introduce  some continual
generalization of usual multiplication of matrices.

  Let $ F_{12} (\Delta (\xi^1); \Theta (\chi^2))$ be a functional of
 functions  $\Delta (\xi)$ and  $\Theta (\chi)$ on two-dimensional
world surfaces $\xi$ and $\chi$ which depends on two fixed points
$\xi^1$ and $\chi^2$.

  Now we consider the functional integral

\begin{eqnarray}
 W(\tilde\Delta(\xi^1); \Delta(\xi^3))=\int D\Theta \exp{iS(\Theta)}
 F_{12}(\tilde\Delta (\xi^1);\Theta(\xi^2)) F_{23} (\Theta(\xi^2);
\Delta(\xi^3))
\end{eqnarray}

( Where $S(\Theta)$ is an action for string fields
  $\Theta(\xi)$)

  as a continual generalization of multiplication of two matrices \\
 $W_{n_{1}n_{3}}=\sum_{n_2} F_{n_{1}n_{2}}F_{n_{2}n_{3}}$.

   Correspondingly we consider the following  multiple functional integral
\begin{eqnarray}
 CTr  F_{12} F_{23} F_{34}\ldots F_{N1}=  \int D\Psi_1
\exp {iS(\Psi_1)} \int D\Psi_2  \exp {iS(\Psi_2)} \ldots \nonumber \\
\int D\Psi_{N}\exp {iS(\Psi_{N})}
F_{12}(\Psi_1; \Psi_2 )
F_{23}(\Psi_2; \Psi_3 ) F_{34}(\Psi_3; \Psi_4 ) \ldots F_{N1}(\Psi_{N};
 \Psi_1 )
\end{eqnarray}
    as a continual generalization  of the trace of the product of N
 matrices.\\

    N-particles string amplitudes for \cite {7} are presented by the
usual functional integral $\int Dg DX DH \exp {iS(X,H,g)}$ and a
continual trace $CTr$ of the product of vertices $V_{i,i+1}$:
\begin{eqnarray}
 A_{N}= \int d^2\xi^1 \int d^2\xi^2\ldots \int d^2\xi^{N}
U(\xi^1,\xi^2 \ldots ,\xi^{N})\nonumber \\ \int DgDXDH \exp
{iS(X,H,g)} CTr \prod_{i}V_{i,i+1}
\end{eqnarray}
\begin{eqnarray}
CTr \prod_{i} V_{i, i+1}=
\int D\Psi_1
D\Phi_1\exp{iS(\Psi_1,\Phi_1)}\;\;\;\;\;\;\;\;\;\;\;\;\nonumber \\ \int
D\Psi_2 D\Phi_2 \exp {iS(\Psi_2,\Phi_2)}\ldots \int D\Psi_{N} D\Phi_{N}
\exp {iS(\Psi_{N},\Phi_{N})} \nonumber \\
V_{12}(\Psi_1(\xi^1),\Phi_1(\xi^1);X(\xi^1),H(\xi^1);
\Psi_2(\xi^1),\Phi_2(\xi^1)) \nonumber \\
V_{23}(\Psi_2(\xi^2),\Phi_2(\xi^2);X(\xi^2),H(\xi^2);
\Psi_3(\xi^2),\Phi_3(\xi^2))
 \ldots              \nonumber \\
V_{N1}(\Psi_{N}(\xi^{N}),\Phi_{N}(\xi^{N});
X(\xi^{N}),H(\xi^N);\Psi_1(\xi^{N}),\Phi_1(\xi^{N}))
\end{eqnarray}

   Vertices of our model are choosing to have a simple
symmetry:
\begin{eqnarray}
V_{ij}(\Psi_i,\Phi_i;X,H;\Psi_j,\Phi_j)=
V_{ij}(\Psi_j,\Phi_j;X,H;\Psi_i,\Phi_i)
\end{eqnarray}

  From here on we shall consider tree  string amplitudes in the
 framework of the operator approach which allow us to obtain
the necessary consistent description of superconformal constraints and
 corresponding vertex operators which satisfy these constraints.
  As usually tree amplitudes of interaction of N strings are
represented by  some integrals of vacuum expectation values of
the product of corresponding operator vertices:
\begin{eqnarray}
  A_{N}=\int \prod^{N}_{i=1} dz_{i} Tr\langle
0|V_{12}V_{23}...V_{N1}|0\rangle
\end{eqnarray}
 For superstring approach operator vertices have the form of commutators
with Virasoro supergenerators:
\begin{eqnarray}
V_{i,i+1}(z_{i})=z_{i}^{-L_{0}}
 \left[ G_{r},W_{i,i+1}(1) \exp{i\tilde p_{i,i+1}X(1)} \right]
z_{i}^{L_{0}}
\end{eqnarray}
\begin{eqnarray}
 \exp{i p_{i,i+1}X(1)}=\exp{i p_{i,i+1}X^{(+)}(1)}
\exp{ik_{i}Y_{i0}} \\ \nonumber
\exp{iq_{i,i+1}X_{0}}\exp{(-ik_{i+1}Y_{i+1,0})}\exp{ip_{i,i+1}X^{(-)}(1)}
\end{eqnarray}
  The momentum of the physical state
 $p_{i,i+1}$ is separated into two parts: the momentum $k_{i}-k_{i+1}$
which corresponds to the new set of two-dimensional fields and
corresponding surfaces and the momentum $q_{i,i+1}$. The last moment
 corresponds to usual two-dimensional NS-surface. In the sector of open
strings:
\begin{eqnarray}
p_{i,i+1}=(k_{i}-k_{i+1})+q_{i,i+1}; \\  \nonumber \;\;
 q_{i,i+1}=\beta (k_{i}+k_{i+1})\\
 k_{i}^2=k_{i+1}^2=\nu^2 ;  \sum_{i=1}^{N} k_{i}=0
\end{eqnarray}
\begin{eqnarray}
\beta^2\sim\frac {\alpha^{\prime}_P}{\alpha^{\prime}_H}
\end{eqnarray}
We have for arbitrary channels in tree
   amplitudes :
\begin{eqnarray}
p_{i,j}=\sum_{l=i}^{l=j}p_{l,l+1}=
 (k_{i}-k_{j})+ \sum_{l=i}^{l=j}q_{l,l+1};
\end{eqnarray}

   In our model we consider factorized vertices in
 correspondence with (3,4).  \begin{eqnarray} W_{i,i+1}(1)= \widetilde
 F_i(\Psi;\Phi) \Pi^{(SF)}F_{i+1}(\Psi;\Phi); i=2,3,...N-1; i\neq 1,N
 \\ W_{1,2}(1)=\widetilde F_1(\Psi';\Phi')F_2(\Psi;\Phi);
W_{N,1}(1)=\widetilde F_N(\Psi;\Phi)F_1(\Psi';\Phi')
\end{eqnarray}

 $\Pi^{(SF)}=|0^{(\Psi\Phi)}\rangle\langle 0^{(\Psi\Phi)}|$
 is the projector onto zero occupation numbers of
 modes of $\Psi$ and $\Phi$ fields.
 It is evidently that namely the projector $\Pi^{(SF)}$ and
separation of $\Psi'$; $\Phi'$ and $\Psi$; $\Phi$-modes bring
amplitudes (5) with vertices (6,7,8) in correspondence with (3,4).
 Now we consider the construction of the Virasoro generators of
 the superconformal algebra for this model.  \\

    The operators $G_{r}$ are  presented  by
 the sum
\begin{eqnarray}
 G_{r}= G_{r}^{NS} + G_{r}^{SF}+G_{r}^{SF'}
\end{eqnarray}
 where
\begin{eqnarray}
G_{r}^{NS}=\frac{1}{2\pi}\int_0^{2\pi} d\tau (H^{\mu}\frac{d}{d\tau}
X_{\mu} + \hat{P}_{\nu}H^{\nu})e^{-ir\tau}
\end{eqnarray}
is the supergenerator to be formed in the ordinary way with help of
the old Neveu-Schwarz fields $X_{\mu}(\tau)$ and $H_{\mu}(\tau)$   \\
$\hat{P}_{\nu}=\sum_{i}\omega_{i\nu}\Gamma_{i}$  \\
for $\nu=0,1,2,3  \;\;\;\;$
$\hat{P}_{\nu}=\Gamma^{(\alpha')} P_{\nu}$; \\
for $\nu=4,5,6,7,8,9 \;\;\;\;  $
$\hat{P}_{\nu}=
(\frac{n_{\nu}}{R_{\nu}}+\sum_{i}\tilde\omega_{i\nu}\Gamma^{cm}_{i\nu})$\\
$n_{\nu}=0;\pm 1;\pm 2;\pm 3 ...$
  and
\begin{eqnarray}
 G_{r}^{SF}= G_{r}^{(0)SF} + \Delta_{r}^{SF}
\end{eqnarray}
 is the supergenerator
 to be formed with help of the new fermion fields $\Psi$ and $\Phi$,
\begin{eqnarray}
G_{r}^{(0)SF}=\frac{1}{2\pi}\int_0^{2\pi}d\tau
\eta(\frac {1}{4}\tilde\Psi\Gamma_{a}\Phi^{a}\Psi+
\sum_{a,b,c}\Phi_{a}\Phi_{b}\Phi_{c})e^{-ir\tau}  \\
\eta=\frac {1}{2i\sqrt{31}}; \\
\Delta_{r}^{SF}=
\frac{1}{2\pi}\int_0^{2\pi}d\tau
\rho^{cd}\tilde\Gamma_{c}\Phi_{d}e^{-ir\tau}
\end{eqnarray}

  $\tilde\Psi=\Psi T_{0}$;\\
 $T_{0}=\gamma_{0}\otimes\tau_{2}\otimes\tau_{2}\otimes\tau_{2}$

  The fields $\Phi_{a},\Phi_{b},\Phi_{c}$  in second sum
$\sum_{a,b,c}\Phi_{a}\Phi_{b}\Phi_{c}$  correspond to
currents $\tilde\Psi\Gamma_{a}\Psi,\tilde\Psi\Gamma_{b}\Psi$ and
 $\tilde\Psi\Gamma_{c}\Psi $ in
the sum $\tilde\Psi\Gamma_{a}\Phi^{a}\Psi$. These matrices
$\Gamma_{a},\Gamma_{b}$ and $\Gamma_{c}$ obey the following equations:
$$
  \Gamma_{a}\Gamma_{a}=\Gamma_{b}\Gamma_{b}=\Gamma_{c}\Gamma_{c}=I
$$
$$
  \left[ \Gamma_{a}, \Gamma_{b} \right]=i\Gamma_{c}
$$

   The supergenerator $G_{r}^{SF'}$ is precisely the same as
 $G_{r}^{SF}$ with substitution of $\Psi'$ and $\Phi'$ fields for
 $\Psi$ and $\Phi$ ones.\\

    The important distiguishing feature  of this approach from previous
works \cite {7,8} is an inclusion of special $32\otimes
32$ matrices $\Gamma$   in the linear parts  $\Delta_{r}^{SF}$ of
$G_{r}^{SF}$ and in the linear part of $G_{r}^{NS}$.  It reminds
introduction of $\gamma_{5}$ and $\gamma_{\mu}$ in the supergenerators
$F_{n}$ of the Ramond model \cite {15}.  So vertex operators become
 matrix operators and then corresponding amplitudes are able to
 reproduce the structure of the dual quark Harari-Rosner diagrams and
the Chan-Paton factor \cite {16} in the natural way.  The matrix
$\Gamma^{(\alpha')}$ provides a separation of hadron and nonhadron
 Regge trajectories in their slopes $\alpha'$.  To be more precise we
suppose some universale slope for hadron trajectories and some very
small slope for nonhadron Regge trajectories in the Born approximation.

    The field $\Phi_{d}$ corresponds to the matrix
$\Gamma_{d}$ which is present in the term
$\tilde\Psi\Gamma_{d}\Phi^{d}\Psi$ of the expression for $G_{r}^{SF}$.
  In its turn $\Phi_{d}$ defines components of the current $ J_{n}^{d}$
  :  $$ J_{n}^{d} = \left[G_{r}^{SF},\Phi^{d}_{n-r} \right].  $$

     For fields of the model under consideration we have the usual
commutation relations corresponding to free two-dimensional fields.
  The commutators for the $X_{\mu}$ components
  $X_{\mu}=\sum_{\atop n\neq 0}\frac
{1}{in}\,a_{n\mu}e^{-in\tau}$ are  \\

  $\left[a_{n\mu},a_{m\nu}\right]=-ng_{\mu\nu}\delta_{n,-m}$;  \\
$\left[P_{\nu},X_{0\mu}\right]=-\frac{1}{i}g_{\mu\nu}$  ;
 $a^{+}_{n\mu}=a_{-n\mu}$  ; \\
$a_{n}|0>=<0|a_{-n}=0$  ;  $n>0$ \\

      The anticommutators  for the others
 $H_{\mu},\Psi_{\alpha\beta\gamma \delta}, \Phi_{a}$
are  following ones: \\

  $H_{\mu}=\sum_{r}b_{r\mu}e^{-ir\tau}$;
  $\Phi_{a}=\sum_{r}\phi_{a,r}e^{-ir\tau}$;

  $\Psi_{\alpha}=\sum_{r}\psi_{\alpha,r}e^{-ir\tau}$;
  $r=\pm 1/2;\pm 3/2;\pm 5/2...$  ; \\
 $\left\{ b_{r\mu},b_{s\nu} \right\}=-g_{\mu\nu}\delta_{r,-s}$;
 $\left\{ \phi_{a,r},\phi_{b,s}\right\}=\delta_{a,b}\delta_{r,-s}$;
  $\left\{\tilde\psi_{\alpha,r},\psi_{\beta,s}\right\}=
 \delta_{\alpha,\beta}\delta_{r,-s}$   \\

    As usually
 $b_{r}^{+}= b_{-r}$  ; $ \phi_{a,r}^{+}= \phi_{a,-r}$ ;
$\psi_{\alpha,r}^{+}= \psi_{\alpha,-r}$\\
    $b_{r}|0>=\phi_{a,r}|0>=\psi_{\alpha,r}|0>=<0|b_{-r}=<0|\phi_{a,-r}=
           <0|\psi_{\alpha,-r}=0$ ; $r>0$  \\

      The anticommutators  for the
 $\Psi'_{\alpha\beta\gamma \delta}, \Phi'_{a}$
are  the same as for $\Psi_{\alpha\beta\gamma \delta}, \Phi_{a}$. \\
   We suppose
 $\left[\tilde\Gamma_{c}, \tilde\Gamma_{c'}\right]=0$;
 $\left[\tilde J_{0}^{c}, \tilde J_{0}^{c'}\right]=0$;

 $\left[\tilde\Gamma_{d}, \hat{P}_{\nu}\right]=0$;
 $\left[\hat{P}_{\mu}, \hat{P}_{\nu}\right]=0$; \\

  Then we get the canonical  superconformal algebra of generators

 $G_{r}=G_{r}^{NS}+G_{r}^{SF}+G_{r}^{SF'}$ :   \\

  $\left\{G_{r},G_{s}\right\}= 2L_{r+s}+\frac{c}{3}(r^2-1/4)\delta_{r,-s}$

   $\left[L_{n},L_{m}\right]=(n-m)L_{n+m}+\frac{c}{12}n(n^2-1)\delta_{n,-m}$

   $\left[L_{n},G_{r}\right]=(n/2-r)G_{n+r}$  \\

  Here $c=3d/2+d'(d'+1)/4;\\  d=10; d'=2^{d/2}=32$  \\

     Due to these commutation relations  additional matrix components
 $\tilde\Gamma_{d}^2$ arise in the operator $L_{0}$ from linear parts
$\tilde\Gamma_{d} \Phi^{d}$ of operators $G^{SF}$. They do not vanish on
 vacuum states and hence will break down  the crossing symmetry and
duality of string amplitudes. We regain  these simmetries by way of a
natural generalization of the matrix parts  of operators
$G_{r}$.
    We shall modify somewhat matrix terms of
$G_{r}$ in order for the operator $L_{0}$ to be vanished  while acting
on the vacuum state. With this aim in mind we introduce operators to be
similar to zeroth modes of the fields  $\Psi$ . Here we do not
consider relation between these modes and  a symmetry of the compact
space of six-dimensional torus. We hope to discuss this point in
future.

   Namely we introduce  usual  spinor operators with
quantum numbers of $\Psi$-fields: \\ $\lambda^{(\pm)}$;
 $\mu^{(\pm)}$;.

  These operators obey simple equations:

$ \left\{\tilde\lambda_{\alpha}^{(-)},\lambda_{\beta}^{(+)}\right\}=
      \delta_{\alpha,\beta}$;
  $\tilde\lambda=\lambda T_{0}$;\\

$ \langle 0|\lambda^{(+)}=0 $; $\lambda^{(-)} |0\rangle=0 $; \\
And we have precisely the same equations for $\mu$ - operators.

  We substitute
  the expressions $\sum_{c,d}\rho_{cd}\tilde\Gamma_{c}\Phi^{d}_r$
  by

$$
\sum_{c,d}\rho_{cd}(\tilde\lambda^{(+)}
\tilde\Gamma_{c}\lambda^{(-)})
\Phi^{d}_r
$$
  and
  the expressions $\sum_{c,d}\rho_{cd}\tilde\Gamma_{c}\Phi'^{d}_r$ in
 $G_{r}^{SF'}$ by

$$
\sum_{c,d}\rho_{cd}(\tilde\mu^{(+)}\tilde\Gamma_{c}\mu^{(-)})
\Phi'^{d}_r
$$

  Besides  we  substitute other  matrix terms in $G_{r}$.

   Namely we substitute the expression
$\hat{P}_{\nu}=\sum_{i}\omega_{i\nu} \Gamma_{i}$ by the
sum \begin{eqnarray} \hat{P}_{\nu}=
\tilde\lambda^{(+)}\Gamma^{(\alpha')}\lambda^{(-)}
\frac{1}{i}\frac \partial {\partial Y_{\lambda0}} +
 \tilde\mu^{(+)}\Gamma^{(\alpha')}\mu^{(-)}
\frac{1}{i}\frac \partial {\partial Y_{\mu0}} +
  \frac{1}{i}\frac \partial {\partial X_{0}}
\end{eqnarray}
 for  $ \nu=0,1,2,3$
\begin{eqnarray}
\Gamma^{(\alpha')}=b_{q}\Pi_{q}+ b_{l}\Pi_{l}
\end{eqnarray}
 and
 by the sum
$$
\frac{n_{\nu}}{R_{\nu}}+\tilde\lambda^{(+)}\Gamma_{\nu}^{cm}\lambda^{(-)}+
\tilde\mu^{(+)}\Gamma_{\nu}^{cm}\mu^{(-)}+J_{\nu}^{cm}+J_{\nu}^{'cm};
$$
$$
n_{\nu}=0,\pm 1,\pm 2,\pm 3,...
$$
for  $ \nu=4,5,6,7,8,9$.
\begin{eqnarray}
 J_{\nu}^{cm} =\frac{2}{\eta}\left[G_{r}^{SF},(\Phi_{\nu}^{cm})_{-r}
\right]
\end{eqnarray}
\begin{eqnarray}
J_{\nu}^{'cm}=\frac{2}{\eta}\left[G_{r}^{SF'},(\Phi_{\nu}^{'cm})_{-r}
\right] \end{eqnarray}

   The field $\Phi_{\nu}^{cm}$ corresponds to the matrix
$\tilde\Gamma_{\nu}^{cm}$.
   Here
$$ \Pi_{q}= I \otimes (\frac{3}{4}I\otimes
I\otimes I - \frac{1}{4}(\tau_{3}\otimes \tau_{3}\otimes I +
\tau_{3}\otimes I\otimes \tau_{3} +I\otimes \tau_{3}\otimes \tau_{3}))
$$
$\Pi_{q}$ is the projector on quark  components \\

$$ \Pi_{l}= I \otimes \frac{1}{4}(I\otimes I\otimes
 I + \tau_{3}\otimes \tau_{3}\otimes I + \tau_{3}\otimes I\otimes
 \tau_{3} +I\otimes \tau_{3}\otimes \tau_{3})
$$
$\Pi_{l}$ is the
 projector on lepton components.

  The currents $J_{\nu}^{cm},
\tilde\lambda^{(+)}\Gamma_{\nu}^{cm}\lambda^{(-)}$;
$\nu=4,5,6,7,8,9$ enter along with  compact components of
momenta $P_\nu$ and related  to quantum numbers of
physical states.\\

  This modification does not change the superconformal algebra of
operators $G_{r}$ considered above and the conditions of nilpotency of
the BRST charge but provides for vanishing of $L_{0}$ on vacuum states,
crossing symmetry and permits to bring  string
amplitudes of our model in accordance with the  quark dual diagrams.

  It is worth to note that the operator $L_0$ may be represented as the
following sum:
\begin{eqnarray}
 L_0=R-\frac{\hat{P}^2}{2} + \Delta^{(L_0)} \;\;\;\;\\
 \Delta^{(L_0)}=\left\{G_{r}^{(0)SF} ,\Delta_{-r}^{SF}\right\}+
 \left\{G_{r}^{(0)SF'} ,\Delta_{-r}^{SF'}\right\}
\end{eqnarray}
 in correspondence with equations (15)-(20).

   Let us introduce some local multiplication $\lambda-$ and
$\mu-$ currents:
$$ \tilde\lambda^{(+)}\Gamma_{i}\lambda^{(-)}*
\tilde\lambda^{(+)}\Gamma_{j}\lambda^{(-)} \equiv
\tilde\lambda^{(+)}\Gamma_{i}\Gamma_{j}\lambda^{(-)}
$$
$$ \tilde\mu^{(+)}\Gamma_{i}\mu^{(-)}*
\tilde\mu^{(+)}\Gamma_{j}\mu^{(-)} \equiv
\tilde\mu^{(+)}\Gamma_{i}\Gamma_{j}\mu^{(-)}
$$

 These results can be obtained as  corresponding operator products:
$$
  \tilde \lambda^{(+)}\Gamma_{i}\Pi^{(\lambda)}\lambda^{(-)}
\tilde\lambda^{(+)}\Gamma_{j}\Pi^{(\lambda)}\lambda^{(-)} \equiv
\tilde\lambda^{(+)}\Gamma_{i}\Gamma_{j}\Pi^{(\lambda)}\lambda^{(-)}
$$
   where
$$
  \Pi^{(\lambda)}\equiv |0^{(\lambda)}\rangle \langle 0^{(\lambda)}|
$$
  is the projector onto zero occupation numbers of the $\lambda$
modes.  And we can wright down similar expressions for $\mu$- modes.

   Now it is possible to formulate our meson amplitudes $A_{N}$ by
entering $\lambda,\mu$-modes in operator vertices (7,8,13,14) and
replacing the operator $\Pi^{(SF)}$ by the product
$\Pi^{(SF)}\Pi^{(\lambda)}\Pi^{(\mu)}$ in the expression (13).
The cyclic operator trace for $\lambda$ and $\mu$-modes
generalizes the Chan-Paton factor in a natural way.
For example  let us consider  the following expression:
$$ T_{N}= \langle
0|\tilde \mu^{(-)}W_{1}\lambda^{(-)} \tilde
\lambda^{(+)}W_{2}\lambda^{(-)} \tilde
\lambda^{(+)}W_{3}\lambda^{(-)}\dots
$$
$$
\tilde\lambda^{(+)}W_{N-1}\lambda^{(-)}
\tilde\lambda^{(+)}W_{N}\mu^{(+)}|0\rangle
$$
   This product turns into the traditional Chan-Paton factor
in the simplest case of
matrices $\Gamma_{i}\Pi^{(\lambda)}$ instead of $W_{i}$:
$$ \langle 0|\tilde
\mu^{(-)}\Gamma_{1}\lambda^{(-)} \tilde
\lambda^{(+)}\Gamma_{2}\Pi^{(\lambda)}\lambda^{(-)}
\tilde\lambda^{(+)}\Gamma_{3}\Pi^{(\lambda)}\lambda^{(-)}\dots
$$
$$
\tilde \lambda^{(+)}\Gamma_{N-1}\Pi^{(\lambda)}\lambda^{(-)}
\tilde \lambda^{(+)}\Gamma_{N}\mu^{(+)}|0\rangle \equiv
Tr (\Gamma_{1}\Gamma_{2}\Gamma_{3}\dots \Gamma_{N})
$$

\section{Nilpotency of the $BRST$-charge operator and  the supercurrent
conditions}

   So we have the additional central charge $c=d'(d'+1)/2$
in the Virasoro superalgebra of $G_{r}$ and $ L_{n} $ operators owing
to the new fields $\Psi$ and $\Phi$.
There is a need to have new additional gauge symmetries
in order that one can
compensate this new central charge by the corresponding central
charge of $BRST$-ghosts and obtain the nilpotency of the $BRST$-charge
operator.

  There are some indications of existence of such symmetries in
appearence of a great many currents to be conserving on the mass shell
in the Bardakci-Halpern string model  \cite {14}. Moreover there are
evidences for these symmetries in the very approach under
consideration.
   Indeed it turns out that the operators $L_{n}^{SF}$ in this model
may be written in Sugavara form  \cite {17}  as
 $L_{n}^{SF}=:\sum_{a,l}J_{l}^{a}J_{l}^{a}:$
 where  $J_{l}^{a}=\left\{G_{l-r},\phi_{r}^{a}\right\}$.
  In general case the Sugavara form to be a normal product of current
constituents contains nonsingular four-fermion components after
its expansion in normal products in relation to fermion constituents.
However here as in  \cite {7,8}  all these four-fermion components are
cancelled and  the operators $L_{n}^{SF}$  acquire  usual form
of the Neveu-Schwarz  model  operators.  The scalar product
  $\sum_{a}J^{a}(\tau)J^{a}(\tau)$  has a symmetry in  relation to
rotations of vectors  $J$. These rotations are generated by some linear
combination $\hat J$ of the very currents.  By virtue of
the commutation relations:\\
$\left[J_{l}^{a},J_{n}^{b}\right]=f_{abc}J_{l+n}^{c}+
l\delta_{l,-n}\delta_ {a,b}$,\\
 invariance of the operator expression
  $\sum_{a}J^{a}(\tau)J^{a}(\tau)$ with respect to these rotations will
  take place in the case that second anomalous term in the commutation
  relations  will be absent for this combination $\hat J$
i.e.\\
  $\left[\hat {J}_{n}^{s},\hat {J}_{-n}^{s}\right]=0$.

   This situation reminds  two-dimensional models with four-fermion
interaction \cite {18}. These models may obey similar symmetries  and
in this case ones have the spectrum of physical states of a free model.
It corresponds to our commutation relations which are valid for free
fields.  On other hand  these new fields $\Psi$ and $\Phi$ lead to new
states of negative norm ( "ghosts").  So the additional symmetries
are necessary for new "ghost" states to be vanished as unitarity
requires.  Due to the superconformal symmetry of our model new current
  gauge conditions to be generated by the operators $\hat J$ must be
accompanied by fermion conditions to be generated by fields $\hat
{\Phi}$ which are superpartners of $\hat J$:
$-l\hat{\Phi}_{r}=\left\{G_{r-l},\hat {J}_{l}\right\}$.

    It is easy to find the number N of necessary additional supercurrent
constraints from the nilpotency BRST-charge $\Omega$: $\Omega^2=0$.
    We have previous BRST-ghost fields of the dimensionality j to be
equal 2 and 3/2 for $L_{n}$ and $G_{r}$ correspondingly. Now new
 BRST-ghost fields of the dimensionality j to be equal  1  and 1/2  for
N new $\hat {J}_{n}^{(i)}$ and  $\hat {\Phi}_{r}^{(i)}$ ($i=1\dots,
N$) constraints have to be added. The contribution of the BRST-ghost
field of the dimensionality j to  central charge  is equal to
$\pm(3(2j-1)^2-1)$.  The sign is determined by the statistics of the
BRST-ghost field. Let us denote the number of boson matter string
  fields by D and the number of fermion fields by $D^{\prime}$. Then we
obtain first condition of the nilpotency BRST-charge  from  the
requirement of the total central charge to be vanished:
$D+\frac{D'}{2}-26+11-2N-N=0$ \\

  For $D=10, D^{\prime}=10+32+32*31/2$  we get  N=88.
 In the case of the arbitrary even dimensionality of D we have

$$
     D^{\prime}=D + 2^{D/2}(2^{D/2}+1)/2
$$

here  $2^{D/2}$ is  the number of spinor components of $\Psi_{\alpha}$;
      $2^{D/2}(2^{D/2}-1)/2$ is  the number of currents from these
  spinor components.
 This condition gives possible values of D:
 $D=6$(mod(4));  $D=6,10,14,18,22,26,...$.

   Second condition of the nilpotency $\Omega^2=0$ requires $L_{0}=1/2$
on physical states. Due to the additional terms
$(\sum_{c,d}\rho_{cd}(\tilde\lambda^{(+)}\tilde\Gamma_{c}\lambda^{(-)})^2$
and  \\
$(\sum_{c,d}\rho_{cd}(\tilde\mu^{(+)}\tilde\Gamma_{c}\mu^{(-)})^2$
in $L_{0}$ this condition does not lead
now to $J=1+\alpha'M^2$  for leading hadron Regge trajectory  and  does
not require  $M^2_{\rho}=0$ correspondingly.

   Third nontrivial condition  from $\Omega^2=0$ results in norms of
the $\hat {\Phi}$ to be vanished. And as consequence the norms
of the currents $\hat J$ are to  be vanished too. In its
turn it brings to the requirement of the vanishing corresponding
Killing metrics $g_{(ij)}$ for group space of $\hat J$:

 $\left[ \hat {J}_{l}^{s},\hat{J}_{-l}^{s}\right]=0$,
 $\left\{ \hat{\Phi}_{r}^{s},\hat{\Phi}_{-r}^{s}\right\}=0$\\

 $g_{(ij)}=\sum_{kl} f_{ikl}f_{jlk}=0$ ;
 $\left[ \hat {J}^{i}_{l},\hat {J}^{j}_{n} \right]=
 f^{ijk}\hat {J}^{k}_{l+n}$  \\

   Now we shall  define current operators $\hat {J}^{(s)}$ for our
nilpotent supercurrent conditions on the basis of general requirements
found above.
  In order to explain the proposed choice of $\hat J$ we shall consider
more simple case \cite {19}:
 $d=6$; the field $\Psi_{\alpha\beta}$ is an eight-component spinor,
 $\alpha=1,2,3,4$ is an usual Dirac index,
 $\beta=1,2$  is an isotopic index.

   We take the field $\Phi_{k}$ to be corresponding
 to the matrix:

  $\Gamma^{k}=\gamma_{\mu}k^{\mu}/\sqrt{k^2}\otimes I=
 \gamma_{0}\otimes I$

  And then we can choose  the required fields  $\hat {\Phi}$  for
supercurrent conditions as the fields to be determined by the matrices
 $\Gamma_ {i}^{k}$:
$$
\Gamma_ {i}^{k}=(I+\Gamma^{k})\Gamma_ {i}
$$
 here  $\Gamma_ {i}$ are matrices which anticommute  with
$\Gamma^{k}$:\\

 $\Gamma_ {i}= \gamma_{5}\otimes \tau_i$; $i=1,2,3$;

$\Gamma_ {j}= \gamma_{j}\otimes I$ ; $j=1,2,3$ ;

 $\left\{ \Gamma_{i,j},\Gamma^{k}\right\}=0$
  We suppose here  $\gamma_{\mu}k^{\mu}/\sqrt{k^2}=\gamma_0$.

 Matrices $\Gamma^{k}\Gamma_ {i,j}$  anticommute  with $\Gamma^{P}$
too. But they do not bring to new matrices $\Gamma_ {i}^{k}$.

   Corresponding to $\Gamma_{i}^{k}$ the currents
$J_ {i}^{k}$ transfer the $\Psi$ components with eigenvalue of
 operator $\Gamma^{k}$ to be equal -1 into other  $\Psi$
components with the same eigenvalue of parity.
Components of $\tilde\Psi=\Psi T_{0}\equiv \Psi \gamma_0 \otimes
\tau_2$  have eigenvalues of $\Gamma^{k}$ which are opposite to
eigenvalues of $\Psi$ components by virtue of antisymmetry of the
$\Gamma^{k}$ matrix. ( We use in our approach matrices $\gamma_{\mu}$
in the Majorana representation.) Namely from
$\Gamma^{k}\Psi=\lambda\Psi$ we have $\tilde\Psi\Gamma^{k}= - \lambda
\tilde\Psi$.  In this simple case the number of corresponding
nondiagonal transitions is equal to $N=4*3/2=6$.  Since  $\Gamma_
 {i}^{k} \Gamma_ {i'}^{k}=0 $  commutators of operators $\hat{\Phi}$
and $\hat J$  are vanished and we have\\
$\hat{J}^{i}_{n}=\left\{G_{n-r},\hat {\Phi}^{i}_{r}\right\}, i=1,2...6$ ;
$\left[\hat {J}_{n},\hat {J}_{-n}\right]=0$  \\

  We note that full current operators $J^{i}_{n}$ defined above
contain the $\Phi^{j}\Phi^{k}f_{ijk}$ terms besides
$\tilde\Psi\Gamma_{i}\Psi$ components.\\

  Let us choose operators $\hat J$ in our model in a
similar way. From point of view  of necessary exclusion of "ghost"
components and quantum numbers of observed states the most appropriate
operator which gives eigenvalues for transitions corresponding to
currents $\hat J$ is the parity operator $J_{P}$.
   The operator $\Phi_{P}$ for the current $J_{P}$
corresponds to the matrix $\Gamma_{P}$:
\begin{eqnarray}
 \Gamma_{P}=\frac{\hat{P}}{\mu}\otimes
 I \otimes I \otimes I
 \end{eqnarray}

 Here  $\mu=\sqrt{P^2}$.

   Now we shall choose 66 current operators $\hat J$  by
using eigen values of $\Psi$ quark components in relation to the
 operator $J_{P}=\frac{2}{\eta}\left\{G_{-r},\hat
{(\Phi}_{P})_{r}\right\}$ for corresponding transitions. The field
 $\Phi_{P}$ corresponds to the matrix $\Gamma_{P}$.  The spinor field
  $\Psi$ has 24 components of upper and lower quarks.  From them
12 components have the eigenvalue of $J_{P}$ to be equal +1
and the other 12 components have
the opposite eigenvalues to be equal -1.  By taking transitions between
the latter components we obtain $12*11/2=66$
transitions and 66 corresponding operators $\hat J$ for quarks.
The matrices for these transitions
are analogical to the above ones
\begin{eqnarray}
\Gamma^{i}_{qP}=(I+\Gamma^{(q)}_{P})\Pi_{q}\Gamma^ {iq}  ;
\end{eqnarray}
\begin{eqnarray}
  \left\{ \Gamma^ {iq},\Gamma^{(q)}_{P}\right\}=0
\end{eqnarray}
\begin{eqnarray}
  \Gamma^{(q)}_{P}= \Gamma_{P}\Pi_{q}
\end{eqnarray}

  In the same way  8 lepton components  of $\Psi$ can be divided in relation
to the eigenvalues of $J^{(P)}$into
4 components with the eigen value of $J^{(l)}_{P}$ to be equal +1 and
4 ones with the eigen value to be equal -1. Hence we have here
$4*3/2=6$ operators to be corresponded transitions from components
$\Psi$ with the eigen value of $J^{(l)}_{P}$ to be equal -1 into ones
with the same eigen value and 6 currents $\hat J$.
\begin{eqnarray}
\Gamma^{i}_{lP}=(I+\Gamma^{(l)}_{P})\Pi_{l}\Gamma^ {il}  ;
\end{eqnarray}
\begin{eqnarray}
  \left\{ \Gamma^ {il},\Gamma^{(l)}_{P}\right\}=0
\end{eqnarray}
\begin{eqnarray}
 \Gamma^{(l)}_{P}=\Gamma_{P}\Pi_{l}
\end{eqnarray}

    Now we shall choose  the remaining 16 operators $\hat J$  as
corresponding components  of $\Psi$:
\begin{eqnarray}
\tilde \Psi (I+ \Gamma_{P})=\tilde \Psi (I+\frac{\hat{P}}{\mu}\otimes
 I \otimes I \otimes I)=- (I- \Gamma_{P})\Psi
\end{eqnarray}

    In all, we have chosen  66+6+16=88
operators $\hat J$ required for the supercurrent constraints.

  It is
easy to see that all commutators of the 88 operators of three
sets are vanished and give the vanishing
corresponding Killing metrics $g_{(ij)}$. All these operators have
vanishing norm. Thus all requirements of the nilpotency of the BRST
 charge are satisfied for this choice of supercurrent constraints.
This choice is quite definite and natural but it can not pretend to the
 uniqueness.
  It is interesting to note that the number of operators $\hat J$ is
equal to 88 only for given division of 32 $\Psi$ components into
 24 quark components and 8 lepton ones. Other divisions lead to other
values of the number of operators $\hat J$.

  Similarly we choose  $\hat {J'}_{n}^{i}$-  and $\hat {\Phi'}_{r}^{i}$
-constraints replacing $\Psi ; \Phi; \lambda $- operators by  $ \Psi';
\Phi';\mu$-operators   in the
expressions considered above.

  As discussed above
we are guided by properties of the real physical states and the
requirements of the generalized algebra of gauge constraints i.e. the
Virasoro superalgebra with the nilpotent supercurrent conditions.

  A wave function of physical state is determined by the expression:
\begin{eqnarray}
  |{Phys}\rangle=G_{\frac {1}{2}}|W_{phys}\rangle
\end{eqnarray}

  So our constraints on the wave function of physical state are
expressed by the following equations:
\begin{eqnarray}
( G_{r}^{NS} + G_{r}^{SF}+G_{r}^{SF'})|W_{phys}\rangle=0;\;\;\;\; \\
( L_{n}^{NS} + L_{n}^{SF}+L_{n}^{SF'})|W_{phys}\rangle=0;
r,n>0
\end{eqnarray}
\begin{eqnarray}
\hat {\Phi}_{r}^{i}|W_{phys}\rangle=0; \;\;\;
\hat {\Phi'}_{r}^{i}|W_{phys}\rangle=0;\;\;  i=1,2,3,...72;\;\;
r>0 \;\;\;                                        \\
(I- \Gamma_{P})\Psi_{r}^{j}|W_{phys}\rangle=0;\;\; \;
 (I- \Gamma_{P})\Psi_{r}^{'j}|W_{phys}\rangle=0;\;\;
j=1,2,3,...16;\;\; r>0\;\; \;\\ \hat {J}_{n}^{i}|W_{phys}\rangle=0;
 \;\;\;\;\;\; \hat {J'}_{n}^{i}|W_{phys}\rangle=0;\;\;
 i=1,2,3,...88;\;\; n=0,1,2,... \;\;\;\; \end{eqnarray}
\begin{eqnarray}
( L_{0}^{NS} + L_{0}^{SF}+ L_{0}^{SF'})|W_{phys}\rangle = \frac{1}{2}
|W_{phys}\rangle;
\end{eqnarray}

    It is worth noting that the evidence of the conditions mentioned
above for physical states in tree amplitudes can be obtained with
help of the expansion of unity for given channel as it was made in
 previous version \cite{7} with the inclusion into
consideration of fields operators to be superfluous for given channel.
In doing so it is necessary at first to separate states generated by
nilpotent supercurrent operators, then states generated by Virasoro
supergenerators with $r,n >1/2$, then states generated by superfluous
for given channel operators and at last by Virasoro supergenerators
with r=1/2.  \\ \\ \\

\section {Fermion states and  elimination of the most part of parity
twins from the baryon spectrum}

     The supercurrent conditions (39) confine essentially  choice of
fermion physical states and eliminate the most part
of parity twins from the fermion spectrum  of physical states
due to the projector $(I- \Gamma_{P})$ in  the operator (34).

   It is necessary to find the formulation of proper operator vertices
which satisfy these supercurrent constraints in fermion channels.
 The direct inclusion of the projector on eigen values +1 of the
current $J_{P}$ is not compatible with  analiticity owing to presence
in $J_{P}$ the singular function $\mu=\sqrt{P^2}$. We can to avoid
this obstacle by replacing $\mu=\sqrt{P^2}$  with an
equivalent nonsingular operator expression which does not break down
the Virasoro conditions.  For this purpose we introduce an additional
dependence of some angle $\phi_i$ in operator vertices $V_i(z_i)$
and corresponding integration over these angles
from 0 to $2\pi$. This construction is similar to the operator vertex
for emission of an usual closed string state \cite{21}:
\begin{eqnarray}
V_{i,i+1}(z_{i};\phi_i)=\phi_i^{R-\tilde R } z_{i}^{-L_{0}}
 \left[  G_{r},W_{i,i+1}(1) \exp{i\tilde p_{i,i+1}X(1)} \right]
 z_{i}^{L_{0}}\phi_i^{-R+\tilde R};
\end{eqnarray}

   Here (see (25),(26))
$$
R=L_0+\frac{\hat{P}^2}{2}-\Delta^{(L_0)}
$$
$$
\tilde
R=2J^{SF}_{\mu\nu}J_{NS}^{\mu\nu}+\frac{1}{4}(1+(-1)^{G_{NS}})+1
;\;\;\mu;\nu=0,1,2,3
$$
$$
G_{NS} =\sum_r (b_{\mu})_{-r}(b^{\mu})_{r}
$$

  The operators $\frac{1}{2}(1\pm (-1)^{G_{NS}})$ are the projectors
onto states which have the positive or negative $G_{NS}$-parity.

  $J_{NS}^{\mu\nu}$ is the part of the  angular momentum
in the Minkowski space which determines by the NS fields i.e.
$X;H$-fields;

 $J^{SF}_{\mu\nu}$ is the operator of the angular momentum for the
$\Psi;\Phi;\Psi';\Phi'$-fields  only.\\

   Let us note that the operators $J_{tot}^{\mu\nu}$ and
$J^{SF}_{\mu\nu}$ are commuting with the Lorentz scalar
supergenerators $G_r$, $L_n$. The operators included into vertices
do not break down the supergauge constraints (36)-(41).

   These operators $R$ and  $\tilde R$ have half-integer and integer
eigen values, the operator $R-\tilde R$ has integer eigen values only.

  The integration over $\phi_i$  leads to the operator
$$
\frac {\sin {\pi(R-\tilde R)}}{\pi(R-\tilde R)}
$$
  and hence to the condition:
\begin{eqnarray}
 R=\tilde R;\;\;\;L_0=2J^{SF}_{\mu\nu}J_{NS}^{\mu\nu}+
\frac{1}{4}(1+(-1)^{G_{NS}})-\frac{\hat{P}^2}{2}+\Delta^{(L_0)}+1.
\end{eqnarray}

   Now we are able to introduce the necessary parity projector for
fermion states in the supergauge invariant manner as the operator
$\Pi^{(\Psi)}_P$ and to substitute our vertices
by $\Pi^{(\Psi)}_PV_{i,i+1}(z_{i};\phi_i)\Pi^{(\Psi)}_P$:
\begin{eqnarray}
 \Pi^{(\Psi)}_P=\tilde J_P-(-1)^{G_{NS}}\hat{M};\\
\tilde J_P=\frac{2}{\eta}\left\{G_{-r},\tilde \Phi^{P}_{r}\right\};\\
 \frac{\hat{M}^2}{2}=2J^{SF}_{\mu\nu}J_{NS}^{\mu\nu}+
\frac{1}{4}(1+(-1)^{G_{NS}})+\Delta^{(L_0)}
\end{eqnarray}
 The operator $\tilde\Phi^{P}$ corresponds to the matrix $\tilde\Gamma_P$:
\begin{eqnarray}
 \tilde\Gamma_P=\hat{P}\otimes I \otimes I \otimes I
\end{eqnarray}

  These modified operator vertices satisfy the conditions (39) and
give some possible solution of the problem of the parity
twins.  Here the most of the parity twins  appear to be spurious
gauge states excluded from the spectrum of the physical states.

\section { Linear terms of the supergauge conditions and
fundamental interactions }

   Now we shall define more precisely linear in fields terms
in the operator $G_{r}^{NS}$ and in the operator
$G_{r}^{SF}=G_{r}^{(0)SF}+\Delta_{r}^{SF}$.  Their detailed form
 determines vertex operators for amplitudes of interaction of particles
and our approach in description of fundamental interactions to a great
extent.  Let us note that our
treatment of gauge supercurrent nilpotent conditions and  formulation
of linear terms here differ essentially from previous versions
\cite{7,20,22}.

  We have rigorous commutation conditions for these linear terms
noted above in second section.
 These requirements of commutations give the following
equations for corresponding matrices:
$$
\left[ \tilde\Gamma_{c}, \tilde\Gamma_{c'} \right]=0;
\left[ \tilde\Gamma_{d}, \Gamma^{(\alpha')} \right]=0;
\left[ \tilde\Gamma_{\nu}^{cm}, \tilde\Gamma_{\nu'}^{cm}\right]=0;
\left[ \tilde\Gamma_{d}, \Gamma_{\nu}^{cm}\right]=0;
\left[ \tilde\Gamma_{\nu}^{cm}, \Gamma^{(\alpha')} \right]=0.
$$
These commutation relations are necessary
for conservation of the form of commutators of $G_{r}$. These
constraints and the Lorentz covariance  restrict significantly a choice
of possible matrices $\tilde\Gamma_{c},\Gamma_{\nu}^{cm}$ and fields
$\Phi^{d}$.  We shall determine fields $\Phi^{d}$ by matrices $
\Gamma_{d}$ which enter  in the part $\tilde\Psi\Gamma_{d}\Phi^{d}\Psi$
 of the operator $G_{r}^{SF}$ with $\Phi^{d}$.  We take the
operators $\tilde J_{d}^2$ which appear in $L_0$ to be commuting with
the operator $J_{P}$  so these fields
$\tilde \Phi^{d}$ will be consistent with the new supercurrent
constraints.

  This choice of suitable matrices $\Gamma_{\nu}^{cm}$ provides a
possibility of electromagnetic and weak fundamental interactions
due to vector mesons arrising in nonplanar one-loop amplitudes (closed
string sector) in addition to usual tensor particles (gravitons).
Their generalized ten-dimensional masses are vanished but usual
four-dimensional masses are nonzero for $W_{\pm}$ and Z-bosons.
 Let us remind that
   the momentum  $P=\sum_i p_{i,i+1}$ is separated into two parts:
\begin{eqnarray}
\sum_i p_{i,i+1}=\sum_i(k_{i}-k_{i+1})+\sum_iq_{i,i+1}; \\ \nonumber
\;\;k_{i}^2=k_{i+1}^2=\nu^2; q_{i,i+1}=\beta (k_{i}+k_{i+1});
\beta^2\sim\frac {\alpha^{\prime}_P}{\alpha^{\prime}_H}.
\end{eqnarray}
  The momentum
$\sum_i(k_{i}-k_{i+1})$ corresponds to the new sets of two-dimensional
fields $(\Psi;\Phi)$ on new two-dimensional surfaces. The momentum
$\sum_i q_{i,i+1}$ corresponds to the usual two-dimensional NS-surface.
In the closed string sector $\sum_i (k_{i}-k_{i+1})$ is vanished and
there are here the momentum $P=\sum_i q_{i,i+1}=\beta \sum_i k_{i}$
only.  Hence the Regge trajectory for the closed string sector takes
the following form :
\begin{eqnarray}
\tilde \alpha(0)+\frac{1}{4}\beta^2 K^2; \;\; K=\sum_i k_{i}
\end{eqnarray}
 instead of
\begin{eqnarray}
 \alpha(0)+\frac{1}{2} P^2; \;\; P \approx \sum_i (k_{i}-k_{i+1})
\end{eqnarray}
for the  open hadron string sector.
And the Regge slope $\alpha^{\prime}$ for the closed
string sector  is equal to $ \frac{1}{2}\beta^2 \alpha^{\prime}_H\sim
\frac{1}{2} \alpha^{\prime}_P\sim (10^{19} Gev )^{-2}$ in this approach.

   Let us take corresponding expressions for the operator
$\Delta_r$ and the matrices $\tilde\Gamma_{c}$ and $\Gamma_{\nu}^{cm}$.

\begin{eqnarray}
\Delta_r^{SF}=
(\alpha_{q}\tilde\lambda^{(+)}\Pi_{q}\lambda^{(-)}+
\frac{1}{4\alpha_{q}}\tilde J_{q})\Phi^{qs}_r; \;\;\;
\Delta_r^{SF'}=(\alpha_{q}\tilde\mu^{(+)}\Pi_{q}\mu^{(-)}+
\frac{1}{4\alpha_{q}}\tilde J'_{q})\Phi'^{qs}_r
\end{eqnarray}
$$ \Pi_{q}= I \otimes (\frac{3}{4}I\otimes
I\otimes I - \frac{1}{4}(\tau_{3}\otimes \tau_{3}\otimes I +
\tau_{3}\otimes I\otimes \tau_{3} +I\otimes \tau_{3}\otimes \tau_{3}))
$$

  The fields $\Phi^{qs},\Phi'^{qs}$ are corresponding to the matrices
$\Gamma_{q}$.
$$
\Gamma_{q}=I \otimes (\frac{3}{4}\tau_{3}\otimes
\tau_{3}\otimes \tau_{3} - \frac{1}{4}(\tau_{3}\otimes I \otimes I+
  I \otimes I\otimes \tau_{3} +I\otimes \tau_{3}\otimes I))
$$

  The currents $\tilde J_{q};\tilde J'_{q}$ are defined
by the fields  $\tilde\Phi^{qs},\tilde\Phi'^{qs}$:
$$
 \tilde J_{q} =\frac{2}{\eta}\left[G_{r}^{SF},\tilde\Phi^{qs}_{-r}
\right]
$$
$$
 \tilde J'_{q}
=\frac{2}{\eta}\left[G_{r}^{SF},\tilde\Phi'^{qs}_{-r} \right]
$$

  The fields $\tilde\Phi^{qs},\tilde\Phi'^{qs}$ are
  corresponding to the matrices $\tilde\Gamma_{q}$:
$$
\tilde\Gamma_{q}=\frac{\hat{P}}{\hat{M}} \otimes
(\frac{3}{4} I \otimes I \otimes I -
\frac{1}{4}(\tau_{3}\otimes \tau_{3} \otimes I+
\tau_{3} \otimes I\otimes \tau_{3}
  +I\otimes \tau_{3}\otimes \tau_{3}))
$$

  The value of coefficient $\alpha_{q}$  provides  the
necessary shift of the intercept of the leading $\rho$- trajectory to
the value equal to $\frac{1}{2}$. It is noteworthy that
the value of $\alpha_{q}^2 > \frac{1}{2}$ bring to unacceptable
negative norms of states i.e. to ghosts in the spectrum of physical
states.\\

  Now we  consider linear  terms for operators to be accompaning
 compact components of momentum. In order
to obtain necessary number of lepton and quark types we extend our set
of the $\lambda$ and  $\mu$ - modes correspondingly and introduce a
triple set of $\lambda^a$; $\mu^a$; $a=1,2,3$.

   This choice gives an appropriate equivalent of colour
quark degrees of freedom and a possibility for three generations of
leptons.
    Let us to write out
corresponding linear terms:
 \begin{eqnarray} \sum_{\gamma=4,5,6,7,8,9}
 (\sum_{a=1,2,3}
    \tilde\lambda^{a(+)}\Gamma_{\gamma}^{a}\lambda^{a(-)}+\sum_{a=1,2,3}
\tilde\mu^{a(+)}\Gamma_{\gamma}^{a}\mu^{a(-)}+
J_{\gamma}^{cm}+J_{\gamma}^{'cm})H^{\gamma}
\end{eqnarray}

\begin{eqnarray}
\sum_{a=1,2,3} \tilde\lambda^{a(+)}\Gamma_4^{a}\lambda^{a(-)}=
\sum_{a=1,2,3}\tilde\lambda^{a(+)}\tilde\Gamma_{e} \lambda^{a(-)}
\end{eqnarray}
\begin{eqnarray}
\Gamma_{e} =\frac{e}{2}(I\otimes \tau_3 \otimes \tau_3 \otimes \tau_3 +
\frac{\hat{P}}{\hat{M}}\otimes I \otimes I \otimes I) \\
\tilde\Gamma_{e}=\tilde e \frac {\hat{k}}{2\nu} \otimes \tau_3 \otimes
\tau_3 \otimes \tau_3
\end{eqnarray}
\begin{eqnarray}
 J_{4}^{cm} =\frac{2}{\eta}\left[G_{r}^{SF},\Phi^{4}_{-r} \right]
\end{eqnarray}

    The field $\Phi_{4}$ corresponds to the matrix
$\Gamma_{e}$.

  This choice suggests  the electric charges to be equal to
$\pm \frac{1}{2}$ for $\lambda,\mu$-components and to $\pm 1,0$ for
$\Psi$- components.

\begin{eqnarray}
\sum_{a=1,2,3}  \tilde\lambda^{a(+)}_R
\Gamma_5^{a}\lambda^{a(-)}_L=
 \sum_{a=1,2,3}\tilde g_W\tilde\lambda^{a(+)}_R
\hat{k}\gamma_5
 \otimes \tau_{1}\otimes \tau_{1}\otimes \tau_{1}
\lambda^{a(-)}_L
\end{eqnarray}
\begin{eqnarray}
 J_{5}^{cm}
 =\frac{2}{\eta}\left[G_{r}^{SF},\Phi^{5}_{-r} \right]
\end{eqnarray}

    The operator $J_{5}^{cm}$ is absent.

 The operators $\lambda^{a(-)}_L$ have the left chirality;
 the operators $\tilde\lambda^{a(+)}_R$ have the right
chirality:

\begin{eqnarray}
 \gamma_5 \lambda^{a(-)}_L=\lambda^{a(-)}_L
\end{eqnarray}
\begin{eqnarray}
 \tilde\lambda^{a(+)}_R\gamma_5 =-\tilde\lambda^{a(+)}_R
\end{eqnarray}

\begin{eqnarray}
\sum_{a=1,2,3} \tilde\lambda^{a(+)} \Gamma_6^{a}\lambda^{a(-)}=
\sum_{a=1,2,3} g_Z\tilde\lambda^{a(+)}(a\Pi_q+b\Pi_l)\lambda^{a(-)}
\end{eqnarray}

\begin{eqnarray}
 J_{6}^{cm} = g_Z\left[G_{r}^{SF},\Phi^{6}_{-r} \right]
\end{eqnarray}

    The field $\Phi_{6}$ corresponds to the matrix
$(\tilde a\Pi_q+\tilde b\Pi_l)$.
\begin{eqnarray}
\hat {P_7}=\alpha_7(\sum_{a=1,2,3} \tilde\lambda^{a(+)}
\Gamma_7^{a}\lambda^{a(-)}+J_{7}^{cm}+\sum_{a=1,2,3}
\tilde\mu^{a(+)}\Gamma_7^{a}\mu^{a(-)}+J_{7}^{'cm})
\end{eqnarray}

\begin{eqnarray}
 J_{7}^{cm} = \frac{2}{\eta}\left[G_{r}^{SF},\Phi^{7}_{-r} \right]
\end{eqnarray}

    The field $\Phi_{7}$ corresponds to the matrix
$\tilde \Gamma_7$:

\begin{eqnarray}
 \Gamma_7=
\frac{1}{2}\Pi_3(\frac{\hat{k}}{\nu}
\otimes\tau_3\otimes \tau_3\otimes\tau_3(M_t-M_b)+M_t\otimes I)+
\;\; \\ \nonumber
f_{12}\Pi_{12}+f_{23}\Pi_{23}+f_{13}\Pi_{13}\;\;\;\;
\end{eqnarray}

\begin{eqnarray}
\tilde \Gamma_7=
\frac{1}{2}\Pi_3(\Gamma_{P}(M_t-M_b)+M_t\otimes \tau_3\otimes
\tau_3\otimes \tau_3)+\;\; \\ \nonumber
f_{12}\Pi_{12}+f_{23}\Pi_{23}+f_{13}\Pi_{13}\;\;\;\;
\end{eqnarray}
\begin{eqnarray}
\Pi_3=\frac{1}{4}[I(\otimes I\otimes I\otimes I+
 \tau_3\otimes \tau_3\otimes I)- (\tau_3\otimes I \otimes \tau_3+
I \otimes \tau_3\otimes \tau_3)]
\end{eqnarray}
\begin{eqnarray}
\Pi_{12}=I\otimes (\tau_{+} \otimes \tau_{-} \otimes I+
 \tau_{-} \otimes \tau_{+} \otimes I)
\end{eqnarray}
\begin{eqnarray}
\Pi_{23}=I\otimes (I \otimes \tau_{+} \otimes \tau_{-} +
I \otimes \tau_{-} \otimes \tau_{+} )
\end{eqnarray}
\begin{eqnarray}
\Pi_{13}=I\otimes (\tau_{+} \otimes I \otimes \tau_{-} +
\tau_{-} \otimes I \otimes \tau_{+} )
\end{eqnarray}

\begin{eqnarray}
\sum_{a=1,2,3}\tilde\lambda^{a(+)} \Gamma_8^{a}\lambda^{a(-)}=
\sum_{a=1,2,3}\tilde\lambda^{a(+)}(\rho_{ae}\Pi_e+\rho_{a\nu}\Pi_{\nu})
\lambda^{a(-)}
\end{eqnarray}

 The operator $J_{8}^{cm}$ is absent.

\begin{eqnarray}
   \Pi_e =\frac 1{2}\Pi_l \otimes(I \otimes I\otimes I -\tau_3
\otimes \tau_3 \otimes \tau_3)
\end{eqnarray}

\begin{eqnarray}
   \Pi_{\nu}=\frac 1{2}\Pi_l\otimes(I \otimes I\otimes I +
\tau_3 \otimes \tau_3 \otimes \tau_3)
\end{eqnarray}

  The coefficients $M_i$;$\rho_i$ and $f_{i}$ can
give necessary mass shifts of quark flavours, lepton masses
and coefficients in the Cabibbo-Kobayashi-Maskawa mixing (CKM) matrix.

\section{Wave functions of underlying physical states in this model}

      Let us define some wave functions in the approach suggested
here:

a) for $\pi$-meson

$$
 \langle 0|\exp {ikX_{0}}
(\tilde\mu^{(-)}\Gamma_{\pi_{(i)}}\lambda^{(-)})G_{\frac{1}{2}}
$$
$$\Gamma_{\pi_{+}}=\gamma_5\otimes
\tau_{+}\otimes\tau_{-}\otimes\tau_{-}; \Gamma_{\pi_{-}}= \gamma_5\otimes
\tau_{-}\otimes \tau_ {+} \otimes \tau_{+};
$$
$$
\Gamma_{\pi_{0}}=\Gamma^{(q1)}\gamma_5\otimes
\tau_3 \otimes \tau_3 \otimes \tau_3
$$

  b) for K-meson
$$
 \langle 0|\exp {ikX_{0}} (\tilde\mu^{(-)}
\Gamma_{K_{(i)}}\lambda^{(-)})G_{\frac{1}{2}}
$$
$$
  \Gamma_{K_{(+)}}=
I \otimes (I-\tau_{3})\otimes (I+\tau_ {3})\otimes \tau_{+}/4
$$
$$
  \Gamma_{K_{(-)}}=
I \otimes (I-\tau_{3})\otimes (I+\tau_ {3})\otimes \tau_{-}/4
$$

 c) for nucleons
$$
 \langle 0| \exp {ikX_{0}} (\tilde\mu^{(-)}\Gamma^{(q1)}
\lambda^{(-)})\tilde U(1+\hat{b^t}_{\frac{1}{2}})
\Gamma^{(q1)}\Psi_{\frac {1}{2}}G_{\frac{1}{2}}
$$
     Here $\hat{b^t}=\hat{b}-\frac{(bP)\hat{P}}{M^2}$.
 d) for leptons of i-th generation
$$
 \langle 0| \exp {ikX_{0}} (\tilde\mu^{(-)}_i\Pi_l
\lambda^{(-)})_i\tilde U\Pi_l\Psi_{\frac {1}{2}}G_{\frac{1}{2}}
$$

  The operator $\Gamma^{(q1)}$ is the projector on the quark components
of first generation:

$$
\Gamma^{(q1)}=
\frac{1}{4}I \otimes ((I\otimes I\otimes I+ I\otimes \tau_{3}\otimes
\tau_{3})- (\tau_{3}\otimes\tau_{3}\otimes I + \tau_{3}\otimes
I\otimes\tau_{3}))
$$
    More detailed analysis of physical states and corresponding
amplitudes would be carried in following publications.

 The author would like to thank E.N.Antonov,A.P.Bukhvostov,G.S.Danilov,
L.N.Lipatov,I.T.Dyatlov,M.G.Ryskin,V.Alexeev and other participants of
the Theoretical Departament of PNPI seminars for useful discussions.

\end{document}